\documentclass[longauth]{aa} 

\usepackage{rotating}
\usepackage{graphicx}
\usepackage{amsmath}
\usepackage{longtable}
\usepackage{enumerate}
\usepackage[flushleft]{threeparttable}

\usepackage{txfonts}
\usepackage{color}

\begin{document}

   \title{The Gaia-ESO Survey: chemical signatures of rocky accretion in a young solar-type star \thanks{Based on observations made with the ESO/VLT, at Paranal Observatory, under program 188.B-3002 (The Gaia-ESO Public Spectroscopic Survey)}}


   \author{ L.~Spina\inst{\ref{inst1},\ref{inst2}}, F. Palla\inst{\ref{inst2}}, S. Randich\inst{\ref{inst2}}, G. Sacco\inst{\ref{inst2}}, R. Jeffries\inst{\ref{inst3}}, L. Magrini\inst{\ref{inst2}}, E. Franciosini\inst{\ref{inst2}}, M.~R. Meyer\inst{\ref{inst4}}, G. Tautvai\v{s}ien\.{e}\inst{\ref{inst5}}, G. Gilmore\inst{\ref{inst6}}, E.~J. Alfaro\inst{\ref{inst7}}, C. Allende Prieto\inst{\ref{inst8},\ref{inst9}}, T. Bensby\inst{\ref{inst10}}, A. Bragaglia\inst{\ref{inst11}}, E. Flaccomio\inst{\ref{inst12}}, S.~E. Koposov\inst{\ref{inst6},\ref{inst13}}, A.~C. Lanzafame\inst{\ref{inst14}}, M.~T. Costado\inst{\ref{inst7}}, A. Hourihane\inst{\ref{inst6}}, C. Lardo\inst{\ref{inst15}}, J. Lewis\inst{\ref{inst6}}, L. Monaco\inst{\ref{inst16}}, L. Morbidelli\inst{\ref{inst2}}, S.~G. Sousa\inst{\ref{inst17}}, C.~C. Worley\inst{\ref{inst6}} and S. Zaggia\inst{\ref{inst18}}}

   \institute{Departamento de Astronomia do IAG/USP, Universidade de S\~ao Paulo, Rua do M\~atao 1226, S\~ao Paulo, 05509-900 SP, Brasil - \email{lspina@usp.br}\label{inst1}
   \and INAF - Osservatorio Astrofisico di Arcetri, Largo E. Fermi, 5, 50125, Firenze, Italy\label{inst2}
   \and Astrophysics Group, Research Institute for the Environment, Physical Sciences and Applied Mathematics, Keele University, Keele, Staffordshire, ST5 5BG, United Kingdom\label{inst3}
   \and Institute for Astronomy, ETH Zurich, Wolfgang-Pauli Strasse 27, 8093 Zurich, Switzerland\label{inst4}
   \and Institute of Theoretical Physics and Astronomy, Vilnius University, A. Gostauto 12, 01108, Vilnius, Lithuania\label{inst5}
   \and Institute of Astronomy, University of Cambridge, Madingley Road, Cambridge CB3 0HA, United Kingdom\label{inst6}
   \and Instituto de Astrof\'{i}sica de Andaluc\'{i}a-CSIC, Apdo. 3004, 18080, Granada, Spain\label{inst7}
   \and Instituto de Astrof\'{\i}sica de Canarias, E-38205 La Laguna, Tenerife, Spain\label{inst8}
   \and Universidad de La Laguna, Dept. Astrof\'{\i}sica, E-38206 La Laguna, Tenerife, Spain\label{inst9}
   \and Lund Observatory, Department of Astronomy and Theoretical Physics, Box 43, SE-221 00 Lund, Sweden\label{inst10}
   \and INAF - Osservatorio Astronomico di Bologna, via Ranzani 1, 40127, Bologna, Italy\label{inst11}
   \and INAF - Osservatorio Astronomico di Palermo, Piazza del Parlamento 1, 90134, Palermo, Italy\label{inst12}
   \and Moscow MV Lomonosov State University, Sternberg Astronomical Institute, Moscow 119992, Russia\label{inst13} 
   \and Dipartimento di Fisica e Astronomia, Sezione Astrofisica, Universit\'{a} di Catania, via S. Sofia 78, 95123, Catania, Italy\label{inst14}
   \and Astrophysics Research Institute, Liverpool John Moores University, 146 Brownlow Hill, Liverpool L3 5RF, United Kingdom\label{inst15}
  \and Departamento de Ciencias Fisicas, Universidad Andres Bello, Republica 220, Santiago, Chile\label{inst16}
   \and Instituto de Astrof\'isica e Ci\^encias do Espa\c{c}o, Universidade do Porto, CAUP, Rua das Estrelas, 4150-762 Porto, Portugal\label{inst17}
   \and INAF - Padova Observatory, Vicolo dell'Osservatorio 5, 35122 Padova, Italy\label{inst18}
   }

   \date{}

 \abstract{It is well known that newly formed planetary systems undergo processes of orbital reconfiguration and planetary migration. As a result, planets or protoplanetary objects may accrete onto the central star, being fused and mixed into its external layers. If 
the accreted mass is sufficiently high and 
 the star has a sufficiently thin convective envelope, such events may result in a modification of the chemical composition of the stellar photosphere in an observable way, enhancing it with elements that were abundant in the accreted mass. The recent Gaia-ESO Survey observations of the 10-20 Myr old Gamma Velorum cluster have enabled identifying a star that is significantly enriched in iron with respect to other cluster members. In this Letter we further investigate the abundance pattern of this star, showing that its abundance anomaly is not limited to iron, but is also present in the refractory elements, whose overabundances are correlated with the condensation temperature. This finding strongly supports the hypothesis of a recent accretion of rocky material.}

 

\authorrunning{L. Spina et al.}
\titlerunning{The Gaia-ESO Survey: chemical indications of rocky accretion in a young solar type star}

   \maketitle

\section{Introduction}

Open clusters are groups of stars formed from the same nebula. For this reason, members of the same cluster are expected to share distance, age, 
kinematics, and chemical content. The latter
reflects the initial composition of the cloud that gave birth to the cluster itself. 

Recently, a few cases of chemically anomalous stars have been observed in open clusters; these can be explained through episodes of planet engulfment or the significant accretion of rocky material \citep{Ashwell05, Spina14}. Stars are subject to occasional accretion events during their lifetime through different processes. In
addition to the well-known phase of gaseous accretion that characterizes newly born stars during their first Myr of evolution, infall of planets or planetesimals onto the central star can represent another important mechanism that is driven by orbital decay following interactions with other massive bodies (e.g., \citealt{Weidenschilling96,Kley12}). The most favorable epoch for such events extends approximately from the end of the main protostellar accretion phase to the completion of pre-main-sequence (PMS) contraction,  when newly formed planets are still clearing their orbits of dust and rocky bodies. A major consequence of the ingestion of (rocky) planetary material is the possible enhancement of photospheric metallicity: after penetrating the stellar atmosphere, rocky masses are rapidly dissolved and mixed with the ambient matter. If the accreting star has a thin convective zone (CZ), the polluting planetary material is not too diluted within the predominantly hydrogen gas and can produce a significant increase of the atmospheric metallicity \citep{Laughlin97, Gonzalez06,Schuler11,Mack14}. Such an enhancement might be detected through high-resolution spectroscopic observations.

The Gamma Vel cluster is a young (10-20~Myr; \citealt{Jeffries14}) open cluster that has recently been observed by the Gaia-ESO Survey \citep{Gilmore12, Randich13}. As we will show, the age of this cluster offers the rare opportunity to investigate the impact that accretion events may have on the composition of the stellar atmosphere. 
Recently, \citet{Spina14} determined the cluster mean metallicity $<$[Fe/H]$>$$=-$0.057$\pm$0.018~dex and found that one of the  members, 2MASS~J08095427$-$4721419 (hereafter, $\#$52 as in \citealt{Spina14}) has
an iron abundance 2$\sigma$ greater than that of the cluster. The star also has an IR-excess at 24~$\mu$m  \citep{Hernandez08}
that is due to a long-lived debris disk that may have already formed planetary-mass objects. Assuming that $\#$52 is genuinely metal-richer than other cluster members, \citet{Spina14} argued that the overabundance might be produced by the ingestion of $\sim$60 M$_\oplus$ of rocky material. Here, we further investigate the peculiarity of $\#$52 by comparing its chemical pattern to that of a similar star of the same cluster. We show that the accretion scenario is confirmed by the overabundance of refractory elements found in the atmosphere of $\#$52.

\section{Dataset}
Gaia-ESO observations of Gamma Vel have been described by \citet{Spina14} and \citet{Jeffries14}. In this paper we use UVES data, in particular, the results from the second analysis cycle that are part of the second internal data release of the Gaia-ESO Survey (GES) in July 2014 (GESviDR2).
Our analysis is based on the recommended abundances 
delivered by two working groups (WGs): one in charge of the UVES spectra of FGK type stars (i.e., WG11), and one dealing with the spectra of PMS stars (i.e., WG12).
Details of the analyses performed by these WGs are reported by \citet{Smiljanic14} and \citet{Lanzafame15}. In both working groups the spectrum analysis is carried out by different nodes, whose results are then combined to provide a set of recommended parameters. In addition to the recommended abundances and stellar parameters provided by the GES, in this study we also exploit the abundances produced by different nodes of the consortium so as to explore the reliability of the analysis, as shown below.

\section{Chemical anomaly of $\#$52}

Abundances of a variety of elements (Na, Mg, Al, Si, Ca, Sc, Ti, V, Cr, Mn, Fe, Co, Ni, and Cu) for two members of Gamma Vel, the metal-rich star $\#$52 and 2MASS~J08093304$-$4737066 (hereafter, $\#$45 as in \citealt{Spina14}), have been determined in GESviDR2 by four nodes. Of all the cluster members, only these two stars are sufficiently slow rotators for a detailed chemical analysis, thus individual abundances could not be derived for other stars in Gamma Vel. However, we stress that star $\#$45 has a metallicity similar to that of the other cluster members, hence it is probably representative of the cluster mean composition. 
We compare the chemical abundances derived for the two stars to gain more insight into the chemical pattern of $\#$52. 
The spectra have a similar S/N (127 and 140, respectively), and from them, the following atmospheric parameters have been derived and released in GESviDR2: star $\#$45 has an effective temperature of 5614$\pm$80~K, a surface gravity of 4.22$\pm$0.12~dex, and a microturbulence of 1.85$\pm$0.08~km/s, while $\#$52 has 5864$\pm$57~K, 4.41$\pm$0.11~dex, and 1.89$\pm$0.10~km/s. 

For each element X, we considered the nodes that produced a A(X)$=$log$N_{X}$/log$N_{H}$+12 value for both stars. Then, for each element, we computed the abundance differences between $\#$52 and $\#$45 determined independently by each node: $\Delta$A(X)$=$A(X$)_{\#52}$$-$A(X$)_{\#45}$; these differential abundances are free from the systematics or zero-points that can affect the analysis of each node. The elemental abundance values of $\#$52 recommended by WG12 (A(X)) and the $\Delta$A(X) derived by each node are listed in Table~\ref{GESdata}. The differential abundances provided by the nodes show a satisfactory agreement for many elements, such as Si, Sc, and $<$Ti$>,$ whose standard deviations around the mean are  0.05, 0.03, and 0.06 dex, respectively. However, for other elements such as Na, Ca, or Ni, one measurement appears to strikingly disagree with the others of the same element. 
To provide a single $\Delta$A(X) for each element, we therefore
opted for the median of all the values since it is a robust measure of the central tendency of the differential abundances given for each element. Thus, in the last column of Table 1, we list the median values $\Delta$\~A of all the elements together with the median absolute deviation (MAD). The uncertainties are typically $\lesssim$0.10~dex, with only one exception: that for copper. Its large error (i.e., 0.41~dex) is due to highly discordant determinations provided by two nodes.




Using the IRAF $\it{splot}$ task, we carefully measured the equivalent widths (EWs) of the Cu lines at 5105.52~$\AA$ in the two spectra, obtaining 101 m$\AA$ for $\#$52 and 113     for $\#$45. The expected uncertainties in the measured EWs can be estimated from the \citet{Cayrel88} formula, which predicts an accuracy of 1m$\AA$ for these lines. We note that this formula neglects the uncertainty due to the continuum setting, but this contribution is minimal in a differential analysis if adopting a similar assumption on the continuum location in both the spectra. Based on these EWs, assuming the stellar parameters mentioned above and the analysis tools used by the GES for the DR2 analysis (i.e., atomic parameters by \citealt{Heiter14} and MARCS stellar models; \citealt{Gustafsson08}), we obtained a differential abundance $\Delta$A(Cu)$=$0.00$\pm$0.04~dex. This error takes into account the uncertainties in the atmospheric parameters of the two stars and has been evaluated as follows: we determined the abundance errors for $\#$52 and $\#$45 as in \citet{Magrini13}, then we adopted as final uncertainty the maximum difference of the abundance values considering their error bars. We neglected the uncertainty due to the EW measurements. A similar check has been performed on the zinc lines at 4810.56~$\AA$: their EWs are 92 m$\AA$ for $\#$52 and 93 for $\#$45. This resulted in a differential abundance $\Delta$A(Zn)$=+$0.02$\pm$0.05~dex. As for Cu, the uncertainty on the EWs (i.e., $\sim$1~m$\AA$) are negligible. These differential abundances are listed in Table~\ref{non-GES data}. By repeating the procedure for four Sc~II lines, we verified that our method provides differential abundances that are consistent with those listed in Table~\ref{GESdata}.

\begin{table}
\tiny
\vspace{-0.2cm}
\begin{center}
\begin{threeparttable}
\caption{\label{GESdata} Differences based on Gaia-ESO node analyses.}
\begin{tabular}{c|c|cccc|c} 
\hline\hline 
Element & A(X) & $\Delta$A$_{1}$ & $\Delta$A$_{2}$ & $\Delta$A$_{3}$ & $\Delta$A$_{4}$ & $\Delta$\~A$\pm$MAD \\ \hline
Na I & 6.24 & 0.62 & $-$0.06 & $-$0.16 & ... & $-$0.06$\pm$0.10 \\
Mg I & 7.74 & ... & 0.33 & ... & 0.12 & 0.23$\pm$0.11 \\
Al I & 6.66 & ... &  0.19 & ... & 0.06 & 0.13$\pm$0.07 \\
Si I & 7.57$\pm$0.14 & 0.12 & 0.08 & 0.04 & 0.18 & 0.10$\pm$0.04 \\
Ca I & 6.48$\pm$0.14 & $-$0.26 & 0.20 & ... & 0.19 & 0.19$\pm$0.01 \\
Sc II & 3.41$\pm$0.20 & 0.24 & ... & ... & 0.29 & 0.27$\pm$0.03 \\
Ti I & 5.16$\pm$0.27 & 0.05 & $-$0.01 & 0.13 & 0.07 & \\
Ti II & 5.31$\pm$0.36 & 0.07 & 0.39 & 0.07 & 0.30 & \\
$<$Ti$>$ & & 0.06 & 0.19 & 0.10 & 0.19 & 0.15$\pm$0.05 \\
V I & 4.22$\pm$0.23 & 0.27 & ... & 0.15 & 0.01 & 0.15$\pm$0.12 \\
Cr I & 5.8$\pm$0.10 & 0.13 & $-$0.02 & 0.18 & 0.08 & 0.09$\pm$0.05 \\
Mn I & 5.84$\pm$0.21 & 0.21 & 0.06 & 0.05 & 0.15 & 0.11$\pm$0.05 \\
Fe I & 7.55$\pm$0.12 & 0.08 & 0.37 & 0.12 & 0.22 & 0.17$\pm$0.07 \\
Co I & 5.16$\pm$0.33 & 0.07 & 0.15 & 0.11 & $-$0.09 & 0.09$\pm$0.04 \\
Ni I & 6.28$\pm$0.14 & 0.15 & 0.12 & 0.42 & 0.11 & 0.14$\pm$0.02 \\
Cu I & ... & $-$0.40 & ... & ... & 0.41 & 0.01$\pm$0.41 \\
\hline\hline 
\end{tabular}
\begin{tablenotes}
      \tiny
      \item a) $<$Ti$>$$=$(log $N_{Ti~I}$ + log $N_{Ti~II}$)/2
      \item b) The Gaia-ESO nodes are 1-EPINARBO, 2-LUMBA, 3-Vilnius,
and 4-Arcetri.
    \end{tablenotes}
\end{threeparttable}
\end{center}
\vspace{-0.3cm}
\end{table}

\begin{table}
\tiny
\begin{center}
\caption{\label{non-GES data} Differences based on a direct analysis of the Gaia-ESO spectra}
\begin{tabular}{c|cc|c} 
\hline\hline 
Feature & EW$_{\#52}$ & EW$_{\#45}$ & $\Delta$A \\ \hline
Cu (5105.52 $\AA$) & 101$\pm$1 m$\AA$ & 113$\pm$1 m$\AA$ & 0.00$\pm$0.04 \\
Zn (4810.56 $\AA$) & 92$\pm$1 m$\AA$ & 93$\pm$1 m$\AA$ & +0.02$\pm$0.05 \\
\hline\hline 
\end{tabular}
\end{center}
\vspace{-0.4cm}
\end{table}

\begin{figure}
\vspace{-0.4cm}
\hspace{-0.6cm}
\vspace{-0.2cm}
\includegraphics[width=0.53\textwidth]{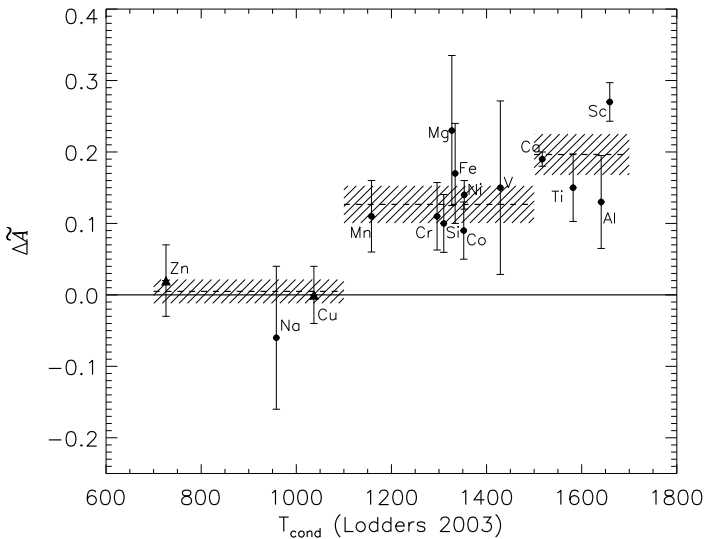}
\caption{Differential abundances of $\#$52 and $\#$45 as a function of the condensation temperature. The plotted $\Delta$\~A values are those reported in Table~\ref{GESdata}. The triangles are from Table \ref{non-GES data}. The condensation temperatures are taken from \citet{Lodders03}. The horizontal dashed lines represent the weighted means of the differential abundances within each class of elements (i.e., volatiles, and moderately and super refractory elements). The bands represent the weighted standard deviation of the mean.}
\label{abu_tcond}
\vspace{-0.3cm}
\end{figure}

The median values $\Delta$\~A are plotted in Fig. \ref{abu_tcond} as a function of the elemental condensation temperature ($T_{cond}$). For the latter we have adopted the values of  \citet{Lodders03}. We also include the values for Cu and Zn from Table \ref{non-GES data} (triangles) that are based on the direct measurements of the lines observed in the two stars. It is possible to separate all the elements plotted in the graph into three illustrative classes: the volatiles would be the elements with $T_{cond}$$<$1100~K (i.e., Zn, Na, and Cu), moderately refractory elements those with 1100$<$$T_{cond}$$<$1500~K (i.e., Mn, Cr, Si, Mg, Fe, Co, and Ni), and the super refractory elements those with $T_{cond}$$>$1500~K (i.e., Ca, Ti, Al, and Sc). For each of these classes we plot in Fig. \ref{abu_tcond} the mean and standard deviation of the differential abundances $\Delta$\~A weighted on the corresponding MADs (dashed lines and oblique lines): 0.00$\pm$0.03~dex for the volatiles, 0.127$\pm$0.014~dex for the moderately refractory and 0.196$\pm$0.009 for the super refractory elements. Thus, even though open clusters are commonly found to be chemically homogeneous (e.g., \citealt{DeSilva06}), the plot shows that all the refractory elements ($T_{cond}$$>$1100~K) are enhanced in $\#$52 by $\gtrsim$0.10~dex. On the other hand, this overabundance is not observed in the volatiles (Zn, Na, and Cu). In addition to this, the super refractory elements are more enhanced than the moderately refractories. Moreover, the most refractory element (Sc) is the most overabundant. Therefore we conclude that the overmetallicity of star $\#$52 with respect to $\#$45 is only limited to the elements with $T_{cond}$$>$1100~K and that the overabundance of a given element is correlated with its condensation temperature.



\section{Enrichment by planet engulfment}
This study represents a significant improvement on the analysis of the anomalous Fe abundance of $\#$52 performed by \citet{Spina14} since it is based on a subsequent internal GES data release with abundances of an extended set of elements obtained by four independent groups. This allows us to place the peculiar chemical abundances on a firmer basis and to investigate now the possible origin of this anomaly, taking advantage of the youth of Gamma Vel, whose age corresponds to the most active and dynamical phase of planet formation and early evolution.

Interestingly, the trend between the overabundance of a given element and its $T_{cond}$ shown in Fig.~\ref{abu_tcond} mirrors the composition seen both in planetary material \citep{Chambers10} and in interstellar dust (e.g., \citealt{Spitzer78}). This suggests that the abundance pattern observed in $\#$52 originated from the enrichment subsequent to the ingestion of planetary-mass-sized rocky objects. Because of their higher condensation temperature, the refractories are thought to be the main components of the solids that accrete onto planets or planetesimals.

Several studies have demonstrated that episodes of planetary material engulfment must be frequent in the highly dynamical environment of young systems. For example, high-precision radial velocity surveys have revealed that about 20-30$\%$ of solar-type stars have low-mass planets (1-30 M$_\oplus$) with orbital periods shorter than 50 days \citep{Mayor08,Howard10}. Several numerical simulations indicate that a large number of terrestrial-planet embryos can form within the innermost regions ($<$1~AU) of a planetary system on timescales shorter than 10~Myr (e.g., \citealt{Chambers01,Ida08,Schlichting14}). Mutual interactions and tidal perturbations of the host star can easily induce their orbital decay and subsequent accretion onto the central star (e.g., \citealt{Zhou10,Ida08,Raymond11}). Similarly, the high frequency of Jupiter-type planets with extremely small orbits clearly points toward efficient planet migration from their birth sites at large distances. During this process, a giant planet can induce other planets to move into unstable orbits \citep{Zhou08}. Such catastrophic events are expected to occur more frequently during the early stages of the evolution of planetary systems (10-100 Myr), when the chaotic growth of planetesimals into planets is still active \citep[e.g.,][and references therein]{Nagasawa07}. 

With an age of $\sim$10-20~Myr, the members of Gamma Vel are prime candidates for studying differential enrichment due to accretion of planetary material. These stars are old enough to host planets or planetesimals that interact with the remnants of the debris disk, but are not too evolved to lose the evidence of the possible modification of their surface abundance. We know that any induced additional metallicity is bound to decrease as a result of several physical processes operating on timescales of several tens to hundreds of Myr. One of the most effective mechanism for the removal of chemical anomalies is the so-called thermohaline convection (e.g., \citealt{Vauclair04}), in which the heavy material accumulated in the external layers of a star produces an unstable weight gradient that triggers the onset of convective instabilities that rapidly sink the overmetallic matter into the radiative stellar interior. Theoretical studies predict that this process would be able to halve the induced overmetallicity in $\sim$50~Myr \citep{Theado12}.

\begin{figure}
\vspace{-0.4cm}
\hspace{-0.5cm}
\includegraphics[width=0.52\textwidth]{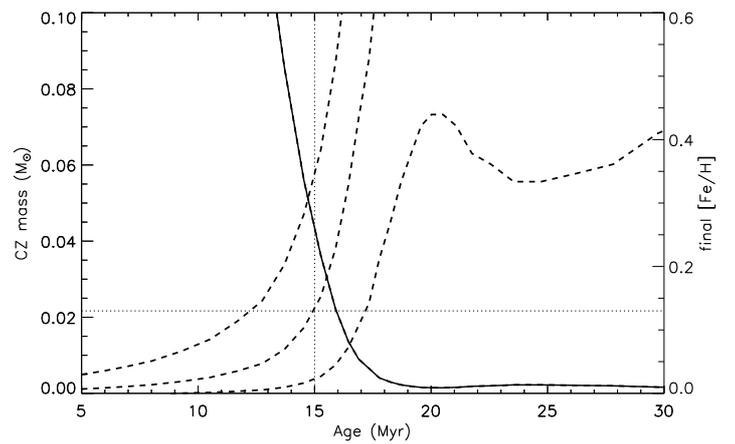}
\caption{Mass contained in the CZ of a 1.3 $M_{\odot}$ star as a function of its age during the PMS contraction phase (solid line). The right scale shows the iron abundance level resulting from accretion. The plot assumes the \citet{Siess00} models for pre-main-sequence stars with solar metallicity. The rise in iron abundance caused by the ingestion of 1, 6, and 20 M$_\oplus$ of pure iron (dashed lines, from right to left) is highly dependent on the thickness of the CZ and, consequently, on the stellar age. The dotted vertical line is located at the age estimated for $\#$52, and the dotted horizontal line refers to the overabundance of $\#$52 relative to Gamma Vel.}
\label{siess_rocky_accr}
\end{figure}

The second aspect to consider is that the mass enclosed in the CZ of a star is a critical parameter for determining the amount of chemical alteration induced by the accreted planetary object. It is well known that, during the early phase of contraction, solar-type and intermediate-mass stars undergo a process of internal readjustment in which the extended inner CZ retreats toward the surface \citep{Palla93}. Using the PMS models of \citet{Siess00}, we illustrate in Fig. \ref{siess_rocky_accr} the variation in CZ mass (solid line) during the first 30~Myr of the evolution of a 1.3~$M_{\odot}$ star, corresponding to the mass estimated for $\#$52 from its $T_{eff}$, V mag, and assuming the Siess tracks. In this case,$\text{ }$the extent of the CZ has shrunk to about 5\% of the total mass  after about 15 Myr and disappears (with
the exception of a thin subphotospheric layer) in another few million years. The figure also shows the time variation of the photospheric iron content of the star (cf. dashed lines) resulting from the accretion of objects with different masses during the PMS contraction phase: 1, 6, and 20 M$_\oplus$ of pure iron. The latter value can be considered as an upper limit that corresponds to the total amount of iron contained in all the planets of our solar system, assuming the elemental abundance distribution of the meteoritic CI chondrites reported by \citet{Lodders03}. As we can see, a fixed accreted mass of rocks can produce extremely different iron enhancements depending on the stellar age owing to the rapidly evolving decrease of the CZ. 

The age of $\#$52 is estimated to be $\sim$15~Myr (vertical dotted line in Fig. \ref{siess_rocky_accr}; \citealt{Spina14}), a value consistent with that of Gamma Vel. From the figure we note that at this evolutionary stage the star has a sufficiently thin CZ that the accretion of 6~M$_\oplus$ of pure iron would be enough to yield the observed iron overabundance (shown by the horizontal dotted line). This amount corresponds to $\sim$30~M$_\oplus$ of rocky material having the chemical composition of CI chondrites \citep{Lodders03}. On the other hand, if $\#$52 were slightly younger (for instance, $\sim$10~Myr), the ingestion of all the iron present in the solar system, would have produced a $\Delta$[Fe/H] $\lesssim$0.08~dex, which is lower than observed. Finally, we see that the ingestion of 1~M$_\oplus$ of pure iron can also cause the observed variation of [Fe/H] if the stellar age is greater than $\sim$17~Myr. Owing to the uncertainty in the stellar age and mass and in the models of PMS evolution, we cannot be sure of the exact values of all the parameters, but the exercise is useful to appreciate that to match the observed overabundance, the accretion episode must have occurred within the last $\sim$5~Myr. 

The PMS models also indicate that, at the current age of Gamma Vel, the solar-type members are in the process of significantly reducing the extent of their CZs to the point where they become thin enough to show some evidence of chemical pollution. Therefore, we may expect that in addition to $\#$52, other stars may have undergone similar episodes of planet engulfment in the past that substantially enriched their chemical composition. However,  such hypothetical metal-rich stars could be rare objects because the accretion events may have occurred more frequently during the earliest stages ($\lesssim$10~Myr) of planetary evolution when the stars still had extended CZs that diluted the chemical contamination. The steep dependence on age of the retreat of the convection zone is another factor that very likely limits the probability of discovering anomalous stars in this or other young clusters. It is also conceivable that other planetary engulfment events might be triggered within the first several hundreds of Myr, during the physical reconfiguration of the planetary system architecture. However, the additional effect of thermohaline convection in reducing or eliminating any sign of enrichment in less than 100 Myr suggests that the only suitable clusters in which to find such chemically peculiar objects are those with ages of between ten and a few hundreds of Myr.

\section{Star $\#$52 in the context of anomalously enriched stars}

Chemical patterns similar to that of $\#$52 have been found in stars of widely different ages. However, different interpretations have been proposed to explain these chemical peculiarities. 

\citet{Melendez09} and \citet{Ramirez09} have found that the Sun is $\it{\text{depleted}}$ of refractory elements with respect to a sample of solar twins. This has been interpreted as evidence of the condensation of refractory elements into planets that occurred in the solar protoplanetary disk, but not around the majority of solar twins. The material locked up into planets did not fall onto the Sun during the main accretion phase and therefore did not pollute the solar CZ. Conversely, the solar
twins that did not form planets were subject to the accretion of the disk material that enhanced the abundance of refractories relative to the Sun. \citet{TucciMaia14} have provided further support to this picture with the observation of a lack of refractory elements in the planet-hosting star 16~Cyg~B with respect to its companion, 16~Cyg~A. The abundance pattern found by Tucci-Maia et al. is remarkably similar to that shown in Fig.~\ref{abu_tcond}. This suggests that the observed chemical anomalies have a similar origin in the formation of rocky bodies. However, we note that the overabundance of refractories found in $\#$52 is considerably larger than that of 16~Cyg~A. As we have shown before, the sensitivity of the
enhancement to the size of the CZ at the time of accretion can account for the observed difference. Considering the age of 16~Cyg~A (7.3~Gyr), other mixing processes may have had time to reduce the initial enhancement.

The pollution from self-enrichment of pure rocky material has also been proposed to explain the overabundance of refractory elements found by \citet{Ashwell05} in the 600~Myr old J37 F-type star with respect to the other members of the NGC~6633 cluster. Notably, the observed enhancement, $\Delta$[Fe/H]$=$0.85, is conspicuous and higher than that found in $\#$52, which has a much thicker CZ. 

Finally, in a sample of 148 solar-type stars \citet{Adibekyan14} have found a clear correlation between abundances and $T_{cond}$ similar to that found in the present study. However, they suggested that this is more likely related to stellar age (e.g., old stars have smaller abundances of refractory elements respect with volatiles) or to Galactocentric distances. Our analysis, based on the comparison of stars of the same cluster, indicates that the peculiar overabundances of refractories found in $\#$52 can be reasonably explained through pollution processes that are due to external causes, such as accretion of rocky material.

\section{Perspectives}

The recent discovery of several stars with a distinctive chemical abundance pattern showing an overabundance of refractory elements over the volatile ones and the correlation with the condensation temperature has opened the observational study of accretion events of rocky bodies (planets, planetesimals) onto their parent stars. The study of the anomalous star $\#$52 of Gamma Vel highlights the importance of studying young systems when the dynamical interaction of forming planets is at a peak and other physical mechanisms operating on longer timescales have not yet had time to wash out possible chemical features. More accurate abundance determinations of both refractory and volatile elements can constrain the nature and the composition of the objects accreted onto the star.
Finally, observations of G- and F-type stars in open clusters with ages of between ten and a few hundreds~Myr will provide the opportunity to investigate the frequency of these chemically polluted stars, the possible causes that trigger the events of planet engulfment (e.g., the presence of a debris disk or a stellar companion), and the fate of the rocky material accreted onto the star.

\begin{acknowledgements}
Based on data products from observations made with ESO telescopes at the La Silla Paranal Observatory under program ID 188.B-3002. These data products have been processed by the Cambridge Astronomy Survey Unit (CASU) at the Institute of Astronomy, University of Cambridge, and by the FLAMES/UVES reduction team at INAF/Osservatorio Astrofisico di Arcetri. These data have been obtained from the Gaia-ESO Survey Data Archive, prepared and hosted by the Wide Field Astronomy Unit, Institute for Astronomy, University of Edinburgh, which is funded by the UK Science and Technology Facilities Council. This work was partly supported by the European Union FP7 programme through ERC grant number 320360 and by the Leverhulme Trust through grant RPG-2012-541. We acknowledge the support from INAF and the Ministero dell' Istruzione, dell' Universit\`a' e della Ricerca (MIUR) in the form of the grant "Premiale VLT 2012" and "The Chemical and Dynamical Evolution of the Milky Way and Local Group Galaxies" (prot. 2010LY5N2T). The results presented here benefit from discussions held during the Gaia-ESO workshops and conferences supported by the ESF (European Science Foundation) through the GREAT Research Network Programme. L.S. acknowledges the support from FAPESP (2014/15706-9) and gratefully thanks J. Mel\'endez for helpful discussions. MRM acknowledges support from the Swiss National Science Foundation (SNF).
\end{acknowledgements}


\bibliographystyle{aa}
\bibliography{/Users/lspina/Copy/papers/bibliography.bib}

\begin{thebibliography}{39}
\expandafter\ifx\csname natexlab\endcsname\relax\def\natexlab#1{#1}\fi

\bibitem[{{Adibekyan} {et~al.}(2014){Adibekyan}, {Gonz{\'a}lez Hern{\'a}ndez},
  {Delgado Mena}, {Sousa}, {Santos}, {Israelian}, {Figueira}, \& {Bertran de
  Lis}}]{Adibekyan14}
{Adibekyan}, V.~Z., {Gonz{\'a}lez Hern{\'a}ndez}, J.~I., {Delgado Mena}, E.,
  {et~al.} 2014, \aap, 564, L15

\bibitem[{{Ashwell} {et~al.}(2005){Ashwell}, {Jeffries}, {Smalley},
  {Deliyannis}, {Steinhauer}, \& {King}}]{Ashwell05}
{Ashwell}, J.~F., {Jeffries}, R.~D., {Smalley}, B., {et~al.} 2005, \mnras, 363,
  L81

\bibitem[{{Cayrel}(1988)}]{Cayrel88}
{Cayrel}, R. 1988, in IAU Symposium, Vol. 132, The Impact of Very High S/N
  Spectroscopy on Stellar Physics, ed. G.~{Cayrel de Strobel} \& M.~{Spite},
  345

\bibitem[{{Chambers}(2001)}]{Chambers01}
{Chambers}, J.~E. 2001, \icarus, 152, 205

\bibitem[{{Chambers}(2010)}]{Chambers10}
{Chambers}, J.~E. 2010, \apj, 724, 92

\bibitem[{{De Silva} {et~al.}(2006){De Silva}, {Sneden}, {Paulson}, {Asplund},
  {Bland-Hawthorn}, {Bessell}, \& {Freeman}}]{DeSilva06}
{De Silva}, G.~M., {Sneden}, C., {Paulson}, D.~B., {et~al.} 2006, \aj, 131, 455

\bibitem[{{Gilmore} {et~al.}(2012){Gilmore}, {Randich}, {Asplund}, {Binney},
  {Bonifacio}, {Drew}, {Feltzing}, {Ferguson}, {Jeffries}, {Micela},
  {Negueruela}, {Prusti}, {Rix}, {Vallenari}, {Alfaro}, {Allende-Prieto},
  {Babusiaux}, {Bensby}, {Blomme}, {Bragaglia}, {Flaccomio}, {Fran{\c c}ois},
  {Irwin}, {Koposov}, {Korn}, {Lanzafame}, {Pancino}, {Paunzen},
  {Recio-Blanco}, {Sacco}, {Smiljanic}, {Van Eck}, \& {Walton}}]{Gilmore12}
{Gilmore}, G., {Randich}, S., {Asplund}, M., {et~al.} 2012, The Messenger, 147,
  25

\bibitem[{{Gonzalez}(2006)}]{Gonzalez06}
{Gonzalez}, G. 2006, \pasp, 118, 1494

\bibitem[{{Gustafsson} {et~al.}(2008){Gustafsson}, {Edvardsson}, {Eriksson},
  {J{\o}rgensen}, {Nordlund}, \& {Plez}}]{Gustafsson08}
{Gustafsson}, B., {Edvardsson}, B., {Eriksson}, K., {et~al.} 2008, \aap, 486,
  951

\bibitem[{Heiter {et~al.}(2014)Heiter, {Gilmore}, {Randich}, \& {et
  al.}}]{Heiter14}
Heiter, {Gilmore}, G., {Randich}, S., \& {et al.} 2014, A\&A, in preparation

\bibitem[{{Hern{\'a}ndez} {et~al.}(2008){Hern{\'a}ndez}, {Hartmann}, {Calvet},
  {Jeffries}, {Gutermuth}, {Muzerolle}, \& {Stauffer}}]{Hernandez08}
{Hern{\'a}ndez}, J., {Hartmann}, L., {Calvet}, N., {et~al.} 2008, \apj, 686,
  1195

\bibitem[{{Howard} {et~al.}(2010){Howard}, {Marcy}, {Johnson}, {Fischer},
  {Wright}, {Isaacson}, {Valenti}, {Anderson}, {Lin}, \& {Ida}}]{Howard10}
{Howard}, A.~W., {Marcy}, G.~W., {Johnson}, J.~A., {et~al.} 2010, Science, 330,
  653

\bibitem[{{Ida} \& {Lin}(2008)}]{Ida08}
{Ida}, S. \& {Lin}, D.~N.~C. 2008, \apj, 673, 487

\bibitem[{{Jeffries} {et~al.}(2014){Jeffries}, {Jackson}, {Cottaar}, {Koposov},
  {Lanzafame}, {Meyer}, {Prisinzano}, {Randich}, {Sacco}, {Brugaletta},
  {Caramazza}, {Damiani}, {Franciosini}, {Frasca}, {Gilmore}, {Feltzing},
  {Micela}, {Alfaro}, {Bensby}, {Pancino}, {Recio-Blanco}, {de Laverny},
  {Lewis}, {Magrini}, {Morbidelli}, {Costado}, {Jofr{\'e}}, {Klutsch}, {Lind},
  \& {Maiorca}}]{Jeffries14}
{Jeffries}, R.~D., {Jackson}, R.~J., {Cottaar}, M., {et~al.} 2014, \aap, 563,
  A94

\bibitem[{{Kley} \& {Nelson}(2012)}]{Kley12}
{Kley}, W. \& {Nelson}, R.~P. 2012, \araa, 50, 211

\bibitem[{{Lanzafame} {et~al.}(2015){Lanzafame}, {Frasca}, {Damiani},
  {Franciosini}, {Cottaar}, {Sousa}, {Tabernero}, {Klutsch}, {Spina}, {Biazzo},
  {Prisinzano}, {Sacco}, {Randich}, {Brugaletta}, {Delgado Mena}, {Adibekyan},
  {Montes}, {Bonito}, {Gameiro}, {Alcal{\'a}}, {Gonz{\'a}lez Hern{\'a}ndez},
  {Jeffries}, {Messina}, {Meyer}, {Gilmore}, {Asplund}, {Binney}, {Bonifacio},
  {Drew}, {Feltzing}, {Ferguson}, {Micela}, {Negueruela}, {Prusti}, {Rix},
  {Vallenari}, {Alfaro}, {Allende Prieto}, {Babusiaux}, {Bensby}, {Blomme},
  {Bragaglia}, {Flaccomio}, {Francois}, {Hambly}, {Irwin}, {Koposov}, {Korn},
  {Smiljanic}, {Van Eck}, {Walton}, {Bayo}, {Bergemann}, {Carraro}, {Costado},
  {Edvardsson}, {Heiter}, {Hill}, {Hourihane}, {Jackson}, {Jofr{\'e}}, {Lardo},
  {Lewis}, {Lind}, {Magrini}, {Marconi}, {Martayan}, {Masseron}, {Monaco},
  {Morbidelli}, {Sbordone}, {Worley}, \& {Zaggia}}]{Lanzafame15}
{Lanzafame}, A.~C., {Frasca}, A., {Damiani}, F., {et~al.} 2015, \aap, 576, A80

\bibitem[{{Laughlin} \& {Adams}(1997)}]{Laughlin97}
{Laughlin}, G. \& {Adams}, F.~C. 1997, \apjl, 491, L51

\bibitem[{{Lodders}(2003)}]{Lodders03}
{Lodders}, K. 2003, \apj, 591, 1220

\bibitem[{{Mack} {et~al.}(2014){Mack}, {Schuler}, {Stassun}, \&
  {Norris}}]{Mack14}
{Mack}, III, C.~E., {Schuler}, S.~C., {Stassun}, K.~G., \& {Norris}, J. 2014,
  \apj, 787, 98

\bibitem[{{Magrini} {et~al.}(2013){Magrini}, {Randich}, {Friel}, {Spina},
  {Jacobson}, {Cantat-Gaudin}, {Donati}, {Baglioni}, {Maiorca}, {Bragaglia},
  {Sordo}, \& {Vallenari}}]{Magrini13}
{Magrini}, L., {Randich}, S., {Friel}, E., {et~al.} 2013, \aap, 558, A38

\bibitem[{{Mayor} \& {Udry}(2008)}]{Mayor08}
{Mayor}, M. \& {Udry}, S. 2008, Physica Scripta Volume T, 130, 014010

\bibitem[{{Mel{\'e}ndez} {et~al.}(2009){Mel{\'e}ndez}, {Asplund}, {Gustafsson},
  \& {Yong}}]{Melendez09}
{Mel{\'e}ndez}, J., {Asplund}, M., {Gustafsson}, B., \& {Yong}, D. 2009, \apjl,
  704, L66

\bibitem[{{Nagasawa} {et~al.}(2007){Nagasawa}, {Thommes}, {Kenyon}, {Bromley},
  \& {Lin}}]{Nagasawa07}
{Nagasawa}, M., {Thommes}, E.~W., {Kenyon}, S.~J., {Bromley}, B.~C., \& {Lin},
  D.~N.~C. 2007, Protostars and Planets V, 639

\bibitem[{{Palla} \& {Stahler}(1993)}]{Palla93}
{Palla}, F. \& {Stahler}, S.~W. 1993, \apj, 418, 414

\bibitem[{{Ram{\'{\i}}rez} {et~al.}(2009){Ram{\'{\i}}rez}, {Mel{\'e}ndez}, \&
  {Asplund}}]{Ramirez09}
{Ram{\'{\i}}rez}, I., {Mel{\'e}ndez}, J., \& {Asplund}, M. 2009, \aap, 508, L17

\bibitem[{{Randich} {et~al.}(2013){Randich}, {Gilmore}, \& {Gaia-ESO
  Consortium}}]{Randich13}
{Randich}, S., {Gilmore}, G., \& {Gaia-ESO Consortium}. 2013, The Messenger,
  154, 47

\bibitem[{{Raymond} {et~al.}(2011){Raymond}, {Armitage}, {Moro-Mart{\'{\i}}n},
  {Booth}, {Wyatt}, {Armstrong}, {Mandell}, {Selsis}, \& {West}}]{Raymond11}
{Raymond}, S.~N., {Armitage}, P.~J., {Moro-Mart{\'{\i}}n}, A., {et~al.} 2011,
  \aap, 530, A62

\bibitem[{{Schlichting}(2014)}]{Schlichting14}
{Schlichting}, H.~E. 2014, \apjl, 795, L15

\bibitem[{{Schuler} {et~al.}(2011){Schuler}, {Flateau}, {Cunha}, {King},
  {Ghezzi}, \& {Smith}}]{Schuler11}
{Schuler}, S.~C., {Flateau}, D., {Cunha}, K., {et~al.} 2011, \apj, 732, 55

\bibitem[{{Siess} {et~al.}(2000){Siess}, {Dufour}, \& {Forestini}}]{Siess00}
{Siess}, L., {Dufour}, E., \& {Forestini}, M. 2000, \aap, 358, 593

\bibitem[{{Smiljanic} {et~al.}(2014){Smiljanic}, {Korn}, {Bergemann}, {Frasca},
  {Magrini}, {Masseron}, {Pancino}, {Ruchti}, {San Roman}, {Sbordone}, {Sousa},
  {Tabernero}, {Tautvai{\v s}ien{\.e}}, {Valentini}, {Weber}, {Worley},
  {Adibekyan}, {Allende Prieto}, {Barisevi{\v c}ius}, {Biazzo},
  {Blanco-Cuaresma}, {Bonifacio}, {Bragaglia}, {Caffau}, {Cantat-Gaudin},
  {Chorniy}, {de Laverny}, {Delgado-Mena}, {Donati}, {Duffau}, {Franciosini},
  {Friel}, {Geisler}, {Gonz{\'a}lez Hern{\'a}ndez}, {Gruyters}, {Guiglion},
  {Hansen}, {Heiter}, {Hill}, {Jacobson}, {Jofre}, {J{\"o}nsson}, {Lanzafame},
  {Lardo}, {Ludwig}, {Maiorca}, {Mikolaitis}, {Montes}, {Morel}, {Mucciarelli},
  {Mu{\~n}oz}, {Nordlander}, {Pasquini}, {Puzeras}, {Recio-Blanco}, {Ryde},
  {Sacco}, {Santos}, {Serenelli}, {Sordo}, {Soubiran}, {Spina}, {Steffen},
  {Vallenari}, {Van Eck}, {Villanova}, {Gilmore}, {Randich}, {Asplund},
  {Binney}, {Drew}, {Feltzing}, {Ferguson}, {Jeffries}, {Micela}, {Negueruela},
  {Prusti}, {Rix}, {Alfaro}, {Babusiaux}, {Bensby}, {Blomme}, {Flaccomio},
  {Fran{\c c}ois}, {Irwin}, {Koposov}, {Walton}, {Bayo}, {Carraro}, {Costado},
  {Damiani}, {Edvardsson}, {Hourihane}, {Jackson}, {Lewis}, {Lind}, {Marconi},
  {Martayan}, {Monaco}, {Morbidelli}, {Prisinzano}, \& {Zaggia}}]{Smiljanic14}
{Smiljanic}, R., {Korn}, A.~J., {Bergemann}, M., {et~al.} 2014, \aap, 570, A122

\bibitem[{{Spina} {et~al.}(2014){Spina}, {Randich}, {Palla}, {Sacco},
  {Magrini}, {Franciosini}, {Morbidelli}, {Prisinzano}, {Alfaro}, {Biazzo},
  {Frasca}, {Gonz{\'a}lez Hern{\'a}ndez}, {Sousa}, {Adibekyan}, {Delgado-Mena},
  {Montes}, {Tabernero}, {Klutsch}, {Gilmore}, {Feltzing}, {Jeffries},
  {Micela}, {Vallenari}, {Bensby}, {Bragaglia}, {Flaccomio}, {Koposov},
  {Lanzafame}, {Pancino}, {Recio-Blanco}, {Smiljanic}, {Costado}, {Damiani},
  {Hill}, {Hourihane}, {Jofr{\'e}}, {de Laverny}, {Masseron}, \&
  {Worley}}]{Spina14}
{Spina}, L., {Randich}, S., {Palla}, F., {et~al.} 2014, \aap, 567, A55

\bibitem[{{Spitzer}(1978)}]{Spitzer78}
{Spitzer}, L. 1978, {Physical processes in the interstellar medium}

\bibitem[{{Th{\'e}ado} \& {Vauclair}(2012)}]{Theado12}
{Th{\'e}ado}, S. \& {Vauclair}, S. 2012, \apj, 744, 123

\bibitem[{{Tucci Maia} {et~al.}(2014){Tucci Maia}, {Mel{\'e}ndez}, \&
  {Ram{\'{\i}}rez}}]{TucciMaia14}
{Tucci Maia}, M., {Mel{\'e}ndez}, J., \& {Ram{\'{\i}}rez}, I. 2014, \apjl, 790,
  L25

\bibitem[{{Vauclair}(2004)}]{Vauclair04}
{Vauclair}, S. 2004, in IAU Symposium, Vol. 224, The A-Star Puzzle, ed.
  J.~{Zverko}, J.~{Ziznovsky}, S.~J. {Adelman}, \& W.~W. {Weiss}, 161--166

\bibitem[{{Weidenschilling} \& {Marzari}(1996)}]{Weidenschilling96}
{Weidenschilling}, S.~J. \& {Marzari}, F. 1996, \nat, 384, 619

\bibitem[{{Zhou}(2010)}]{Zhou10}
{Zhou}, J.-L. 2010, in EAS Publications Series, Vol.~42, EAS Publications
  Series, ed. K.~{Go{\.z}dziewski}, A.~{Niedzielski}, \& J.~{Schneider},
  255--266

\bibitem[{{Zhou} \& {Lin}(2008)}]{Zhou08}
{Zhou}, J.-L. \& {Lin}, D.~N.~C. 2008, in IAU Symposium, Vol. 249, IAU
  Symposium, ed. Y.-S. {Sun}, S.~{Ferraz-Mello}, \& J.-L. {Zhou}, 285--291

\end{thebibliography}

\end{document}